\documentclass[preprint,showpacs,preprintnumbers,amsmath,amssymb]{revtex4}
\usepackage{graphicx}
\usepackage{dcolumn}
\usepackage{bm}

\newcommand{\be}{\begin{equation}}
\newcommand{\en}{\end{equation}}
\newcommand{\bea}{\begin{eqnarray}}
\newcommand{\ena}{\end{eqnarray}}

\begin{document}


\title{ Tachyon warm inflationary universe model in the weak dissipative regime }

\author{Sergio del Campo}
 \affiliation{ Instituto de F\'{\i}sica, Pontificia Universidad
Cat\'{o}lica de Valpara\'{\i}so, Casilla 4059, Valpara\'{\i}so,
Chile.}
\author{Ram\'on Herrera}
 \affiliation{ Instituto de
F\'{\i}sica, Pontificia Universidad Cat\'{o}lica de
Valpara\'{\i}so, Casilla 4059, Valpara\'{\i}so, Chile.}
\author{Joel Saavedra}
 \affiliation{ Instituto de F\'{\i}sica, Pontificia Universidad
Cat\'{o}lica de Valpara\'{\i}so, Casilla 4059, Valpara\'{\i}so,
Chile.}

\begin{abstract}
 Warm inflationary universe model in a tachyon field theory is studied in the weak dissipative regime.
 We develop our model  for an exponential potential and the dissipation parameter $\Gamma=\Gamma_0$=constant.
  We describe scalar  and tensor perturbations for this scenario.
\end{abstract}

\maketitle

\section{Introduction}

It is well known that several long-standing problems of the Big Bang
model (horizon, flatness, monopoles, etc.) may find a natural
solution in the framework  of the inflationary universe model
\cite{guth,infla}. One of the  successes  of the inflationary
universe model is that it provides a causal interpretation of the
origin of the observed anisotropy of the cosmic microwave
background (CMB) radiation and also the distribution of large
scale structures \cite{WMAP}.

Warm inflation is an alternative mechanism for produce successful
inflation and avoiding the reheating period \cite{warm}. In warm
inflation, dissipative effects are important during inflation, so
that radiation production occurs concurrently  with the
inflationary expansion. The dissipating effects arises from a
friction term that describes dissipating the processes of the scalar field
 into a thermal bath via its interaction with other
fields. Also, warm inflation shows how thermal fluctuations during
inflation may  play a dominant role in the production of initial
perturbations. In such models, the density fluctuations arise from
thermal rather than quantum fluctuations \cite{62526}. Among the
most attractive features of these models, warm inflation end  at
the epoch when the universe stops inflating and "smoothly" enters
in a radiation dominated Big-Bang phase\cite{warm}. The matter
components of the universe are created by the decay of either the
remaining inflationary field or the dominant radiation field
\cite{taylorberera}.

On the other hand, implications of string/M-theory to
Friedmann-Robertson-Walker (FRW) cosmological models have been
attracted  great attention in the late time, in particular, those related to
brane-antibrane configurations  such as space-like
branes\cite{sen1}. The tachyon field associated with unstable
D-branes might be responsible for cosmological inflation in the
early evolution of the universe, due to tachyon condensation near
the top of the effective scalar potential  which could also add
some new form of cosmological dark matter at late times
\cite{Sami_taq}.

The outline of the paper is a follows. In  section \ref{secti},
the dynamics of tachyon warm inflationary model is obtained.
In section \ref{pert}, cosmological perturbations are
investigated.  Finally, in section \ref{conclu}, we give some
conclusions.

\section{Tachyon Warm Inflationary Model \label{secti}}

As was pointed by  Gibbons \cite{gibbons},  the energy density,
$\rho_\phi$, and pressure, $p_\phi$, associated with the tachyon
field are defined  by $\rho_\phi=V(\phi)/\sqrt{1-\dot{\phi}^2}$
and $p_\phi=-V(\phi)\,\sqrt{1-\dot{\phi}^2}$, respectively. Here,
$\phi$ denotes the tachyon field (with unit $1/m_p$, where $m_p$
represents the Planck mass ) and $V(\phi)=V$ is the effective
potential associated with this  tachyon field. The  potential is
one that satisfies $V(\phi)\longrightarrow$ 0 as
$\phi\longrightarrow \infty$.  It has been argued  that qualitative tachyonic potential of string theory can be described
via an exponential potential of the form \cite{sen1}
\begin{equation}
V(\phi)=V_0 e^{-\alpha\phi},\label{pot}
\end{equation}
where $\alpha$ and $V_0$ are free parameters. In the following we
will take $\alpha> 0$ (with unit $m_p$). Note that $\alpha$
represents the tachyon mass
\cite{Fairbairn:2002yp,delaMacorra:2006tm}. In Ref.\cite{Sami_taq}
is given an estimation of these parameters  in the limit
$A\rightarrow 0$. Here, it was found  $V_0\sim 10^{-10}m_p^4$ and
$\alpha\sim 10^{-6} m_p$. We should mention here that the caustic
problem  with multi-valued regions for scalar Born-Infeld theories
with an exponential potential results in high order spatial
derivatives of the tachyon field, $\phi$, become
divergent \cite{ko}.

 The dynamics of the FRW cosmological model in the warm inflationary scenario,
is described by the equations
 \be
 H^2\,=\kappa \left[\rho_\phi+\rho_\gamma \right]=
 \kappa \left[\frac{V}{\sqrt{1-\dot{\phi}^2}}+\rho_\gamma \right]\label{key_02},
 \en
 \be
\dot{\rho}_\phi+3\,H\,(\rho_\phi+p_\phi)=-\Gamma\dot{\phi}^2\;\;\Rightarrow\;\,\,
 \frac{\ddot{\phi}}{(1-\dot{\phi}^2)}+\,3H \;
\dot{\phi}+\frac{V_{,\,\phi}}{V}=-\frac{\Gamma}{V}\sqrt{1-\dot{\phi}^2}\;\;\dot{\phi},
\label{key_01}
 \en
and \be \dot{\rho}_\gamma+4H\rho_\gamma=\Gamma\dot{\phi}^2
,\label{3}\en where  $H=\dot{a}/a$ is the Hubble factor, $a$ is a
scale factor, $\rho_\gamma$ is the energy density of the radiation
field and $\Gamma$ is a dissipation coefficient, with unit
$m_p^5$. Dissipative coefficient  is responsible for the decay
of the tachyon scalar field into radiation during the inflationary
regime \cite{Moss,Bastero}. Dissipation coefficient, $\Gamma$
can be assumed as a function of $\phi$ \cite{2}, and  thus
$\Gamma=f(\phi)>0$ by the second law of thermodynamics. Dots mean
derivatives with respect to cosmological time, $V_{,\,\phi}=\partial
V(\phi)/\partial\phi$ and $\kappa=8\pi/(3m_p^2)$.

During the inflationary era the energy density associated with the
tachyonic field is the order of the potential, i.e. $\rho_\phi\sim
V$, and dominates over the energy density associated with the
radiation field, i.e. $\rho_\phi>\rho_\gamma$.

With $\Gamma=\Gamma_0=const.$ and using the exponential potential
given by Eq.(\ref{pot}), we find that the slow roll parameter
become
\begin{equation}
\varepsilon=-\frac{\dot{H}}{H^2}=\frac{1}{6\kappa}
\left[\frac{V_{,\,\phi}}{V}\right]^2\frac{1}{V}=\frac{1}{6\kappa}\frac{\alpha^2}{V_0
e^{-\alpha\phi}}.\label{eta1}
\end{equation}
Assuming the set of slow-roll conditions, $\dot{\phi}^2 \ll 1$,
and $\ddot{\phi}\ll 3H(1+r)\dot{\phi}\sim 3H\dot{\phi}$, the
Hubble parameter is given by $H(\phi)=\sqrt{\kappa V_0}
e^{-\alpha\phi/2}$, where the rate $r$ becomes
\begin{equation}
r=\frac{\Gamma}{3HV}=\frac{m_p\Gamma_0}{\sqrt{24\pi}}\;\frac{1}{V_0^{3/2}}\;
e^{ \,3\alpha\phi/2}< 1,\label{rate}
\end{equation}
and parameterizes the  dissipation of our model. For the weak (or
high) dissipation  regime, $r< 1$ (or $r\gg 1$).

 The evolution of $\dot{\phi}$
during this scenario is governed by the expression
$\dot{\phi}=-V_{,\,\phi}/3  H  V$. In the following, the
subscripts $i$ and $f$ are used to denote the beginning  and the
end of inflation.

Using Eq. (\ref{key_02}), the total number of e-folds  at the end
of warm inflation results as
\begin{equation}
N_{total}=-3\kappa\int_{\phi_i}^{\phi_f}\frac{V^2}{V_{,\,\tilde{\phi}}}\;
d\tilde{\phi}=\frac{3\kappa}{\alpha^2}[V_i-V_f],
\end{equation}
where the initial tachyonic field satisfies  $\phi_i<\phi_f$,
since  $V_i>V_f$. Rewriting  the total number of e-folds  in terms
of  $V_f$ and $V_i$, and  using  that $\varepsilon_f\simeq 1$, we
find  $V_i=(2N_{total}+1)V_f$. Since,  the $N_{total}$ parameter
could assume appropriate values (at least 60)
 in order to solved standards cosmological puzzles.
 To do this, we need the following inequality must be satisfied:
 $V_i>10^2\,V_f$.

\section{The Perturbations  \label{pert}}

In this section we will describe scalar perturbations in the
longitudinal gauge, and then we will continue describing tensor
perturbations.

By using the longitudinal gauge in the perturbed FRW metric, we
write \cite{wa}
\begin{equation}
ds^2=(1+2\Phi)dt^2-a(t)^2(1-2\Psi)\delta_{ij}dx^{i}dx^{j},
\end{equation}
where $\Phi=\Phi(t,{\bf x})$ and $\Psi=\Psi(t,{\bf x})$ are
gauge-invariant variables introduced by Bardeen \cite{Barde}.
Since that we need the non-decreasing adiabatic and
isocurvature modes on large scale $k\ll a H$, (which turn out to
be  weak time dependent quantities), when $k$ is expressed in the momentum space, and combining  with the slow
roll conditions we may define  $\Phi$, $\delta\phi$,
$\delta\rho_\gamma$, and $v$  (we omit the subscript $k$
here) \cite{yo} by following equations
\begin{equation}
\Phi\simeq\frac{4\pi}{m_p^2}\left(\frac{V\dot{\phi}}{H}
\right)\;\left[
1+\frac{\Gamma}{4HV}+\frac{\Gamma_{,\,\phi}\dot{\phi}}{48 H^2 V}
\right]\;\delta\phi,\label{PHI}
\end{equation}
\begin{equation}
\hspace{-1. cm}\left[3H+\frac{\Gamma}{V}\right](\delta\phi\dot{)}+
\left[(\ln(V))_{,\,\phi\phi}+\dot{\phi}\left(\frac{\Gamma}{V}\right)_{,\,\phi}\right]\;\delta\phi\;
 \simeq
\left[\dot{\phi}\frac{\Gamma}{V}-2(\ln(V))_{,\;\phi}
\right]\;\Phi,
\end{equation}
\begin{equation}
\delta\rho_\gamma\simeq\frac{\dot{\phi}^2}{4H}[\Gamma_{,\;\phi}\delta\phi-3
\Gamma\Phi]\Longrightarrow\;\frac{\delta\rho_\gamma}{\rho_\gamma}\simeq\frac{\Gamma_{,\;\phi}}{\Gamma}
\;\delta\phi-3\Phi,
\end{equation}
and
\begin{equation}
 v\simeq
-\frac{k}{4aH}\left[\Phi+\frac{\delta\rho_\gamma}{4\rho_\gamma}+
\frac{3\Gamma\dot{\phi}}{4\rho_\gamma}\;\delta\phi\right].
\end{equation}
Here  $v$ appears  from the decomposition of the
velocity field $\delta u_j =-\frac{i a k_j}{k}\;v\;e^{i {\bf kx}}
$ $(j=1,2,3)$ \cite{Barde}.

Note that in the case of the scalar perturbations tachyon and
radiation fields are interacting. Therefore, isocurvature (or
entropy) perturbations are generated, besides  the adiabatic ones.
This occurs because warm inflation can be considered as an
inflationary model with two basics fields. In this context,
dissipative effects themselves can produce a variety of spectral
ranging between red and blue \cite{62526}, thus producing the
running blue to red spectral as  suggested by WMAP five-year
data\cite{WMAP}.

The above equations can be solved taking $\phi$ as an independent
variable instead of $t$. With the help of Eq. (\ref{key_02}) we
find
$$
\left(3H+\frac{\Gamma}{V}\right)\frac{d}{dt}=\left(3H+\frac{\Gamma}{V}\right)\,\dot{\phi}\frac{d}{d\phi}=
-(\ln(V))_{,\,\phi}\frac{d}{d\phi},
$$
and  introducing an auxiliary function $\varphi$ given by
\begin{equation}
\varphi=\frac{\delta\phi}{(\ln(V))_{,\,\phi}}\exp\left[\int
\frac{1}{(3H+\Gamma/V)}\left(\frac{\Gamma}{V}\right)_{,\,\phi}\;d\phi\right],\label{solvar}
\end{equation}
we  obtain the following equation for $\varphi$
\begin{equation}
\frac{\varphi_{,\,\phi}}{\varphi}=-\frac{9}{8}\frac{(\Gamma/V+2H)}{(\Gamma/V+3H)^2}\;
\left[\Gamma+4HV-\frac{\Gamma_{,\,\phi}(\ln(V))_{,\,\phi}}{12H(3H+\Gamma/V)}\right]
\,\frac{(\ln(V))_{,\,\phi}}{V}.\label{var}
\end{equation}
Solving Eq.(\ref{var}) for $\Gamma=\Gamma_0=$ constant and using
Eq.(\ref{solvar}) and condition $r<1$, we find that
$\delta\phi=C\,(\ln(V))_{,\,\phi} \exp[\Im(\phi)]$, where $C$ is a integration
constant
 and
 \begin{equation}
\Im(\phi)=-\frac{9}{2}\int\;H(\ln(V))_{,\,\phi}\left[\frac{(\Gamma_0/V+2H)}{(\Gamma_0/V+3H)^2}
\right]\;d\,\phi,
\end{equation}
or equivalently
\begin{equation}
\Im(\phi)=-\frac{9}{2}\int\;\frac{H}{V}\,\left[\frac{(\Gamma_0/V+2H)}{(\Gamma_0/V+3H)^2}
\right]\;d\,V.
\end{equation}

In this way, the expression for the density perturbations for
$\Gamma=\Gamma_0=constant$, becomes
\begin{equation}
\delta_H=\frac{2}{5}\frac{m_p^2\,\exp[-\Im(\phi)]}{(\ln(V))_{,\,\phi}}\,\delta\phi.\label{33}
\end{equation}
We noted here that in the case $\Gamma=0$, Eq.(\ref{33}) is
reduced to $\delta_H\sim V\delta\phi/(H\dot{\phi})\sim
H\delta\phi/\dot{\phi}$, which coincides with expression
obtained in cool inflation.


The fluctuations of the tachyon field are generated by thermal
interaction with the radiation field, instead of quantum
fluctuations. Therefore, we may write  in the case $r< 1$, that $
(\delta\phi)^2\simeq\,H\,T_r/2\,m_p^4\,\pi^2$, where $T_r$  is the
temperature of the thermal bath \cite{B1}.


On the other hand, from Eqs.(\ref{key_02}) and (\ref{key_01}),
under slow roll approximations and a quasi-stable state, i.e.
$\dot{\rho}_\gamma\ll\Gamma\dot{\phi}^2$, together with
$\rho_\gamma=\sigma\,T_r^4$, ($\sigma$  the Stefan-Boltzmann
constant) we get
$$
T_r=\left[\frac{\Gamma_0}{36\sigma\,H^3}\;\frac{V'\,^2}{V^2}\right]^{1/4},
$$
and from Eq.(\ref{33}), we find
\begin{equation}
\frac{d\ln\delta_H^{2}}{d\phi}=\left[-2\,\frac{d\Im(\phi)}{d\phi}+\frac{V'}{8\,V}\right].\label{gg}
\end{equation}

 The scalar spectral index $n_s$, is defined  by
\begin{equation}
n_s-1=\frac{d\ln\delta_H^2}{d\ln k},\label{ss}
\end{equation}
where the interval in wave number and the number of e-folds are
related by $d\ln k(\phi)\simeq d\,N(\phi)$. By using Eqs.
(\ref{pot}), (\ref{gg}) and  (\ref{ss}), we get
\begin{equation}
n_s-1\simeq\,-\frac{17\,\alpha^2}{24\,\kappa\,V}.\label{nsa}
\end{equation}
Note that $n_s-1$ is $-\frac{\alpha^2}{24\,\kappa\,V}$ bigger than
that obtained in the tachyonic cold inflation case, where
$n_s-1\approx- 2\alpha^2/(3\kappa V)$ \cite{no33}.  The warm
inflation   expression for $n_s$ in the weak dissipative regimen
for the standard case was done in Ref.\cite{au}.

The rate $r<1$ (characteristic of the weak dissipative regime)
allows
 to establish a condition for the ratio $\Gamma_0/\alpha^3$. This condition in terms of the scalar index is given by
\begin{equation}
\frac{\Gamma_0}{\alpha^3}<3\,\left[\frac{17}{24}\right]^{3/2}\,\frac{1}{\kappa\,(1-ns)^{3/2}}.\label{cons}
\end{equation}
Here, we  have used  Eqs.(\ref{rate}) and (\ref{nsa}). In
particular, for $n_s\simeq 0.97$ the above inequality is well
supported for $\Gamma_0< 42\,\alpha^3\,m_p^2$.

The generation of tensor perturbation during inflation would
produce stimulated emission in the thermal background of the
gravitational wave \cite{yo}. From expression (\ref{33}), we may
write the tensor-scalar ratio $R(k)=(A^2_g/P_{\cal R})$ as \be
R(k_0)=\left.\frac{48\,\pi}{ m_p^2}\,
\left[\frac{\varepsilon\,H^3}{T_r}\,e^{2\,\Im(\phi)}
\,\coth\left(\frac{k}{2T}\right)\right]\right|_{\,k=k_0},\label{R}
 \en
where we have used expressions $\delta_H\equiv 2P_{\cal R}^{1/2}/5$ and
$A_g^2=32V\coth[k/2T]/(3m_p^4)$, whith  $k_0$  is referred to as the
pivot point.

From the combination of WMAP five-year data \cite{WMAP}  with the
Sloan Digital Sky Survey (SDSS) large scale structure surveys
\cite{Teg}, there is found an upper bound $R(k_0$=0.002
Mpc$^{-1}$)$ <0.28 (95\% CL)$, where $k_0$=0.002 Mpc$^{-1}$
corresponds to $l=\tau_0 k_0\simeq 30$, with the distance to the
decoupling surface $\tau_0$= 14400 Mpc. SDSS measures galaxy
distributions at red-shifts $a\sim 0.1$ and probes $k$ in the
range 0.016 $h$ Mpc$^{-1}$$<k<$0.011 $h$ Mpc$^{-1}$. Also WMAP
five-year results gives the values for  the scalar curvature
spectrum $P_{\cal R}(k_0)\equiv\,25\delta_H^2(k_0)/4\simeq
2.4\times\,10^{-9}$ and the scalar-tensor ration $R(k_0)=0.095$.
Using the WMAP five-year data and choosing the parameters
$T\simeq\,T_r\simeq 0.24 \times 10^{16}$ GeV and
$k_0=0.002$Mpc$^{-1}$. We obtained from Eqs.(\ref{33}) and
(\ref{R}), that $V(\phi_0)\sim \, 10^{-12} m_p^4$ and $\alpha\sim
10^{-5}\;m_p$. Also, we would like to note that from
Eq.(\ref{cons}) when the dissipative effects have a negligible
influence ($\Gamma_0<10^{-14}m_p^5$) warm inflation occurs in the
so-called weak dissipative regime for $n_s\simeq 0.97$.

\section{Conclusions \label{conclu}}

In this paper we have investigated the tachyonic warm inflationary
scenario in the weak dissipative  regime. Our  specific model is
described by an exponential scalar potential where the dissipation
coefficient, $\Gamma=\Gamma_0=constant$. In relation to the
corresponding perturbations, we  found a general relation for the
density perturbation expressed by Eq.(\ref{33}). The tensor-scalar
ratio is modified by a temperature dependent factor via stimulated
emission into the existing thermal background (see Eq.(\ref{R})).

Using the WMAP five-year data, we have found some constraints
for the parameters appearing in our model. For example,  the
potential becomes of the order of $V(\phi_0)\sim 10^{-12}m_p^4$
when it leaves the horizon, at the scale of $k_0=0.002$Mpc$^{-1}$,
and the parameter $\alpha\sim 10^{-5}m_p$. From Eq.(\ref{cons}) we
obtained a constrain from the dissipative parameter
$\Gamma=\Gamma_0=const.<42\,\alpha^3 m_p^2$ for $n_s=
 0.97$. In particular,   for $\alpha\sim 10^{-5}m_p$
the  dissipation coefficient
$\Gamma=\Gamma_0=const.<10^{-14}m_p^5$
 in order that weak dissipative regime occurs.

Dissipative effects plays a crucial role in producing the entropy
mode; they  can themselves produce a rich variety of  spectra
ranging between red and blue. The possibility of a spectrum which
runs from blue to red is particularly interesting because it is
not commonly seen in inflationary models, which typically predict
red spectral. Models of inflation with dissipative effects and
models with interacting fields have much more freedom than single self-interacting scalar field
to fit the observational data. Summarizing, we have been successful in described
tachyon warm inflationary model for characterize the early epoch of
the universe in the weak dissipation regime.

\vspace{-.50cm}\begin{acknowledgments}
 S.d.C. was
supported by COMISION NACIONAL DE CIENCIAS Y TECNOLOGIA through
FONDECYT grant N$^0$ 1070306. Also, from UCV-DGIP N$^0$ 123.787
(2008).  R. H. was supported by the ``Programa Bicentenario de
Ciencia y Tecnolog\'{\i}a" through the grant  \mbox {N$^0$
PSD/06}. J. S. was supported by DI-PUCV 123.789-2006 and by
Fondecyt grant N$^0$ 11060515.
\end{acknowledgments}


\end{document}